\documentclass[prb,aps,twocolumn,psfig]{revtex4}
\usepackage{epsfig}
\usepackage{graphicx}
\usepackage{amsmath}

\begin{document}
%\draft

%@twocolumnfalse\endcsname

\title{Dynamical properties of  
liquid Al near melting. An orbital-free molecular dynamics study}

\author{D.J. Gonz\'alez\footnote{e-mail: david@liq1.fam.cie.uva.es}, 
L. E. Gonz\'alez and  J. M. L\'opez} 

\affiliation{Departamento de F\'\i sica Te\'orica, Universidad de Valladolid,
Valladolid, SPAIN}

\author{ M. J. Stott}

\affiliation{ Department of Physics, Queen's University, 
Kingston, Ontario, CANADA}

\date{\today}

\begin{abstract}
The static and dynamic structure of liquid Al is studied using the 
orbital free {\em ab-initio} molecular dynamics method. 
Two thermodynamic states along the coexistence 
line are considered, namely T = 943 K and 1323 K for which X-ray 
and neutron scattering 
data are available. A   
new kinetic energy functional, which fulfills a number of
physically relevant conditions is employed, along with a local first 
principles pseudopotential. 
In addition to a comparison with experiment, we also 
compare our {\em ab-initio} results with those obtained from conventional 
molecular dynamics simulations using effective interionic 
pair potentials derived from second order pseudopotential 
perturbation theory. 
\end{abstract}

\maketitle

%]

%\narrowtext

%\footnotetext{$^*$ e-mail: david@liq1.fam.cie.uva.es}

\section{Introduction.}

Molecular dynamics (MD) methods have a long tradition 
as a useful technique to study the properties of liquid 
systems, and the last fifteen years have witnessed a large spread on 
the application of {\it ab-initio} molecular dynamics 
methods, based on the density 
functional theory (DFT). This theory allows calculation of the ground 
state electronic energy of a collection of atoms, for given nuclear 
positions, \cite{Hoh-Kohn,Kohn-Sham} and also yields the forces on the 
nuclei via the Hellmann-Feynman theorem. It enables to perform  
molecular dynamics simulations where the nuclear positions evolve according to 
classical mechanics whereas the electronic subsystem follows adiabatically.

In this paper we present the results of an {\it ab-initio} molecular 
dynamics simulation on the static and dynamic properties of liquid 
Al at thermodynamic conditions around the triple point.  
Liquid aluminium has usually been considered as a simple metal in 
which the core electrons forming the ion can be clearly distinguished 
from the valence electrons, and moreover the core electrons do not 
significantly overlap with those of neighbouring ions. 
Therefore, the system consists of a binary mixture of ions and valence 
electrons, where the former may be 
treated classically whereas the  electrons must be treated quantum 
mechanically.

However, in  wide regions of the density-temperature
plane, the simple metals have usually been treated as 
an effective one-component fluid of ions interacting by means of  
density-dependent effective interionic pair potentials, derived 
from ionic pseudopotentials by applying 
%linear response theory and 
second order perturbation theory. 
This approach, which  will be refered to as the linear response 
theory (LRT), has often been used as the starting point 
for the study of the static and 
dynamic properties of the simple metals. \cite{Hugh,Ng,Kahl,Canales} 
It has also been the approach followed in most studies 
on the static structure of liquid 
aluminium. \cite{Ebbsjo,DRT,Jacucci,Bretonnet,Hafner&Jank} 
Among them, we mention the work of Dagens {\it et al.}  
\cite {DRT} who obtained an effective interionic pair potential, derived from 
a non-local pseudopotential which was constructed from the valence 
charge density induced by an Al$^{+3}$ ion placed in an electron gas at 
the metallic density. From this potential,  
Jacucci {\it et al.} \cite{Jacucci} calculated the static structure 
factor of liquid aluminium by means of MD simulations; their results 
showed fair agreement with experiment, with a main peak  
somewhat higher.  
Also Hafner and Jank \cite{Hafner&Jank} have studied the liquid static 
structure of aluminium by means of an effective interionic pair potential 
derived 
from an {\it ab-initio} pseudopotential originally developed by Harrison,  
\cite{Harrison} whereas the corresponding liquid static structure was 
derived by means of MD simulations.  
In their calculation of the pseudopotential, the authors used the 
coefficient for the exchange-correlation potential between the core and  
valence electrons as a fitting parameter in order to  
obtain agreement with the experimental static structure factor. 

Whereas the previous work dealt only with the static properties of 
liquid Al near melting, the work by Ebbsjo {\it et al.} \cite{Ebbsjo} also 
considered some dynamic properties. In fact, these authors performed 
MD simulations for three different interionic pair potentials, two of them 
based on non-local pseudopotentials and the other one based on the 
local Ashcroft's pseudopotential, \cite{Ashcroft} which  
showed rather different shape, specially outside the repulsive core. 
Despite those differences, all them gave  
fairly similar results for the liquid static structure  which 
agreed well with the experimental data whereas the main 
discrepancies appeared in the dynamic structure. 

A rather different approach has been followed by Chihara and 
coworkers. \cite {Chihara&Kambayasi,Anta&Louis} Their  
quantum hypernetted chain (QHNC) method treats the 
ions and electrons on basically equal footing by combining 
liquid state integral equations with the density functional 
formalism. Moreover, it does not rely on the  pseudopotential 
ideas, gives rise to 
a self-consistent scheme to determine the liquid static structure and 
yields an effective interionic pair potential which depends on 
the particular liquid static structure.

Although the LRT approach has produced reasonable results for 
the liquid alkali metals, 
when the valence of the system is increased, its validity becomes 
more questionable. In addition, even for the alkali metals the LRT is 
less justifiable  
for thermodynamic states approaching the critical point, 
and it is certainly wrong near the critical point.
This limitation of the ``standard" theory has estimulated the use of
first-principles molecular dynamics techniques,   
\cite{kozo1,kozo2,cabral,kresse,bickham} where the electronic
density, total energy and forces are obtained  by using the 
Kohn-Sham (KS) formulation of the density functional
theory (DFT). \cite{Kohn-Sham}  However, the computational demands 
of these {\em ab-initio} methods, where KS orbitals are used to 
describe the electronic density and 
to compute exactly the electronic kinetic energy,
grow very rapidly with system size,
and their memory requirement is also quite large. 
These considerations have restricted the sizes of the systems 
studied so far, to about 
60 atoms, and have limited  simulation times to around 2-5
picoseconds in the cases of Rb, Cs and Hg,\cite{kozo1,kozo2,cabral,kresse}
and 64-128 atoms, with simulation times of 0.15-0.85 picoseconds in the case of 
Na.\cite{bickham} 
These limitations can be at least partly 
overcome if the
exact calculation of the electronic kinetic energy is given up in favour of an
approximate kinetic energy functional of the electronic density. Within
this scheme \cite{Madden1} the number of variables describing the
electronic state  are enormously reduced, especially for large systems,  
enabling the study of larger systems for longer simulation times.
This approach has already been used for several studies on 
solids, \cite{ofdft1} clusters,\cite{ofdft2} and  
some liquid metals (Li, Na, Mg, Al) near melting.\cite{ofdft3,Anta}
We have recently presented \cite{sara} an application of this method to 
study the static structure and some dynamic properties of 
expanded liquid Cs, for which experimental data 
is available. \cite{Winter1} 
In that study 125 particles were used and the simulation time 
was 17-35 picoseconds after an equilibration time of 11-25 picoseconds.
Also, another study \cite{GGLS1} with 205 particles for liquid Al  
gave results for the static structure in good agreement with experiment.  
Recently, Anta {\it et al.} \cite{Anta} have also applied the same scheme  
to study the ionic and electronic static structure of liquid Al near 
melting, leading to results for the static structure factor in 
excellent agreement with experiment. 

The static structure factor 
of liquid Al has been 
measured by both neutron \cite{Jovic,Takeda} and 
X-ray \cite{Waseda,SQM-LIQ} diffraction. 
The dynamical structure of liquid Al near the triple point has also been 
investigated recently by  
Scopigno {\it et al.} \cite{Scopigno} using inelastic 
X-ray scattering (IXS).  
Note that the high value of the adiabatic sound speed for 
liquid Al ( $\approx 4800$ m/s),  prevents the use of the inelastic 
neutron scattering (INS) technique for investigating the collective 
excitations for small $q$-values (roughly, for $q \leq q_p$, 
with $q_p \approx 2.70$ \AA$^{-1}$ being the main peak 
position of the static structure factor). Those IXS experiments 
have investigated the wavevector region 
0.05$q_p \leq q \leq $ 0.5$q_p$, obtaining several dynamical features 
previously observed in the liquid alkali metals, such as the existence 
of collective excitations up to $q$-values larger that 0.5$q_p$, which exhibit 
a positive dispersion in the sound velocity with respect to the 
hydrodynamic value.

The layout of the paper is as follows. In section \ref{theory} we briefly
describe the theory used in the orbital-free {\em ab-initio} 
molecular dynamics (OF-AIMD) simulations, giving some technical details,
and focusing on the two problematic issues, namely, the kinetic 
energy functional and 
the local pseudopotentials needed to characterize the 
ion-electron interaction.  
In section \ref{results} we present and discuss 
the results of the {\em ab-initio} simulations; moreover they are 
compared with further classical molecular dynamics (CMD) simulations 
that we have performed based on LRT and the QHNC potentials,   
and with the available experimental data. 
Finally some conclusions are drawn and possible ideas for further 
improvements are suggested.

\section{Theory.}
\label{theory}

The total potential energy of a system of $N$ classical ions enclosed in a 
volume $V$, and interacting with 
$N_{\rm e}=NZ$ valence electrons through a local
electron-ion potential $v(r)$, is
written, within the Born-Oppenheimer approximation, as the sum of the
direct ion-ion coulombic interaction energy, and the
ground state energy of the electronic system subject to the external
potential created by the ions, $V_{\rm ext}
(\vec{r},\{\vec{R}_l\}) = \sum_{i=1}^N v(|\vec{r}-\vec{R}_i|)$ ,

\begin{equation}
E(\{\vec{R}_l\}) = \sum_{i<j} \frac{Z^2}{|\vec{R}_i-\vec{R}_j|} +
E_g[\rho_g(\vec{r}),V_{\rm ext}(\vec{r},\{\vec{R}_l\})] \, ,
\end{equation}

\noindent where $\rho_g(\vec{r})$ is the ground state electronic density and 
$\vec{R}_l$ are the ionic positions. 
According to LRT, the ground state electronic density is given, 
in reciprocal space, by

\begin{equation}
\rho_g^{\rm LRT} (\vec{q}) = 
\left( \sum_j e^{i\vec{q}\vec{R}_j} \right) n^{\rm LRT}(q)
\equiv F(\vec{q})\; n^{\rm LRT}(q)
\label{LRT1} 
\end{equation}
\begin{equation}
n^{\rm LRT}(q) = \chi(q,\rho_0) v(q) \label{LRT2}
\end{equation}

\noindent
where $\chi(q,\rho_0)$ is the response function of a uniform electron gas
of density $\rho_0 = NZ/V$. 
Accordingly, the ground state electronic density is a
superposition of spherically symmetric pseudoatomic densities around each
ion, i.e.,

\begin{equation}
\rho_g^{\rm LRT} (\vec{r}) = 
\sum_j n^{\rm LRT}\left(\left|\vec{r}-\vec{R}_j\right|\right)
\end{equation}

\noindent
and the electronic ground state energy is  

\begin{eqnarray}
E_g^{\rm LRT} = E_v[\rho_0] + \sum_{i<j} \phi_{\rm ind}(R_{ij}) \\
\phi_{\rm ind}(q) = \chi(q,\rho_0) v^2(q)
\label{phind}
\end{eqnarray}
where $E_v[\rho_0]$ is a structure-independent term. 
Within the LRT, the total potential energy can be written as a sum
of a structure-independent term and a sum over pairs of an effective interionic 
pair potential $\phi_{\rm eff}(R) = Z^2/R + \phi_{\rm ind}(R)$.

Alternatively, DFT shows that the ground state electronic density can
be obtained by minimizing the energy functional
$E[\rho]$, and the minimum value of the functional gives the ground state 
energy of the electronic system. 
The energy functional can be written   

\begin{equation}
E[\rho(\vec{r})] = 
T_s[\rho]+ E_{\rm ext}[\rho]+ E_H[\rho]+ E_{\rm xc}[\rho]
\label{etotal}
\end{equation}

\noindent
where the terms represent, respectively, 
the electronic kinetic energy, $T_s[\rho]$, 
of a non-interacting system of density $\rho(\vec{r})$, 
the energy of
interaction with the external potential due to the ions, 

\begin{equation}
E_{\rm ext}[\rho] = \int d\vec{r} \, \rho(\vec{r}) V_{\rm ext}(\vec{r}) \, ,
\end{equation}
the classical electrostatic energy (Hartree term), 

\begin{equation}
E_H[\rho] = \frac12 \int \int d\vec{r} \, 
d\vec{s} \, \frac{\rho(\vec{r})\rho(\vec{s})}
{|\vec{r}-\vec{s}|} \, ,
\end{equation}
and the exchange-correlation
energy, $E_{\rm xc}[\rho]$, for which we will adopt the local 
density approximation.

\subsection{Technical details}

Given an explicit functional $T_s[\rho]$, we can proceed to minimize 
$E_g$ with respect to the $\rho(\vec{r})$, but in order to maintain 
$\rho ( \vec{r}) \geq 0$ everywhere, we have used as our system 
variable, an effective orbital, $\psi(\vec{r})$, defined as 
$\rho(\vec{r})=\psi(\vec{r})^2$, with 
real $\psi(\vec{r})$. 
We expand $\psi(\vec{r})$ in plane waves 
compatible with the simple cubic periodic boundary conditions 
of the simulation:

\begin{eqnarray}
\psi(\vec{r})=\sum_{\vec{G}}c_{\vec{G}} \; e^{-i\vec{G}\cdot\vec{r}} \\
c_{\vec{G}} = \frac{1}{V}\int_Vd\vec{r}\,\psi(\vec{r}) \; e^{i\vec{G}\cdot\vec{r}}
\\
\vec{G}=\frac{2\pi}{L} \; (n_1,n_2,n_3) \, .
\end{eqnarray}

\noindent where $L$ stands for the side of the cube. 
This expansion is truncated at wavevectors corresponding to a given
cut-off energy, $E_{\rm Cut}$, whose value is given in Table \ref{states}. 
A real $\psi$ implies that $c_{-\vec{G}}=c_{\vec{G}}^*$, with a real 
$c_0$; consequently only the half-set $\{c_{\vec{G}}\}'$s need be treated 
as variables. 

The energy functional must be minimized with the normalization constraint 
${\cal G}[\rho(\vec{r})] = \int_Vd\vec{r}\rho(\vec{r}) = N_{\rm e}$   
which is imposed via the Lagrange multiplier $\mu$, leading to the 
Euler-Lagrange equation

\begin{equation}
\frac{\delta {\cal F}}{\delta\rho(\vec{r})} \equiv 
\frac{\delta [E-\mu{\cal G}]}{\delta\rho(\vec{r})} =
\frac{\delta E}{\delta\rho(\vec{r})} - \mu \equiv \mu(\vec{r}) - \mu = 0
\label{eulereq}
\end{equation}

\noindent for the ground state density.
The minimization is performed with respect to the  
$\{c_{\vec{G}}\}'$s, instead of the electronic density, leading 
to the equations:

\begin{eqnarray}
\frac{\partial{\cal F}}{\partial c_0}  = 
2\int_Vd\vec{r}\,\mu(\vec{r})\psi(\vec{r}) - 2 \mu V c_0 = 0 \nonumber \\
\frac{\partial{\cal F}}{\partial c_{\vec{G}}}  = 
4\int_Vd\vec{r}\,\mu(\vec{r})\psi(\vec{r})e^{i\vec{G}\cdot\vec{r}} - 
4 \mu V c_{\vec{G}} = 0 
\label{gradient}
\end{eqnarray}

\noindent for the ground state density. The minimization of the
functional is performed every time step of the simulation, using a simple
quenching method: a fictitious ``coefficients' kinetic energy", 
${\cal T} = \frac{1}{2} M_c \sum_{\vec{G}} |\dot{c}_{\vec{G}}|^2$, is 
introduced, where $M_c$ is the "coefficients' mass", and the dot denotes
the derivative with respect to the fictitious ``coefficients' time", $t_c$. 
This kinetic energy, rewritten in terms of the set $\{c_{\vec{G}}\}$, 
together with the ``potential energy" ${\cal F}$, leads to the following 
``equations of motion"
($\forall c_{\vec{G}} \in \{c_{\vec{G}}\}$)

\begin{equation}
M_c \ddot{c}_{\vec{G}} = -2 \int_V d\vec{r} \, 
\mu(\vec{r})\psi(\vec{r})e^{i\vec{G}\cdot\vec{r}} + 2 \mu V c_{\vec{G}} 
\,\,\,\,\, 
\label{driv_force}
\end{equation}

These equations are solved numerically using the Verlet leapfrog algorithm
\cite{leapfrog} with an electronic  timestep $\Delta t_c$.
The velocities are quenched at every step 
until the minimum is reached within preset tolerances on 
${\cal T}$ and the  
gradient of ${\cal F}$.   
The chemical potential $\mu$ is not known in advance of the minimization, but 
replacing $\mu$ in eqn. (\ref{driv_force}) by its stationary 
value $\int d \vec{r} \mu(\vec{r}) n(\vec{r}) /  \int d \vec{r}  n(\vec{r})$ 
at each timestep, gives good convergence to the ground state.   
For the present simulations, we have used 
$M_c=1.85\times 10^7$ hartree $\times$ (a.u.)$^3$ and a 
$\Delta t_c=1\times 10^{-4}$ ps.

The interatomic forces are obtained from the electronic ground state 
via the Hellman-Feynman theorem,
$\vec{F}_i=-\vec{\nabla}_{\vec{R}_i}E_g[\rho(\vec{r}),\{\vec{R}_l\}]$,
 ( $i$ = $1 \cdots N$) and Newton's 
equations, $d^2\vec{R}_i/dt^2=\vec{F}_i/M_i$, are solved 
numerically for the motion of the ions using the  Verlet leapfrog algorithm 
with a timestep $\Delta$$t$ = 1.5 $\times 10^{-3}$ ps.

\subsection{The kinetic energy functional}

The kinetic energy functional, $T_s$, is a critical ingredient 
of the energy functional. 
It is generally considered 
\cite{kryachko}
that the von Weizs\"acker term,

\begin{equation}
T_W[\rho(\vec{r})] = \frac18 \int d\vec{r} \, |\nabla \rho(\vec{r})|^2
/\rho(\vec{r}), 
\end{equation}

\noindent is essential for a good description of the kinetic energy. 
It applies in the case of rapidly varying densities, and it
is exact for one or two-electron systems. 
Further terms are usually added to the
functional in order to reproduce correctly some exactly known limits. In
the uniform density limit, the exact kinetic energy is given
by the Thomas-Fermi functional, 

\begin{equation}
T_{\rm TF}[\rho(\vec{r})] = \frac{3}{10} \int
d\vec{r} \, \rho(\vec{r}) k_F(\vec{r})^2, 
\end{equation}

\noindent where $k_F(\vec{r})=(3\pi^2)^{1/3}\rho(\vec{r})^{1/3}$ is the
local Fermi wavevector. 
In the limit of almost uniform density, LRT is
correct, with a response function corresponding to a non-interacting
uniform electron gas, given by the Lindhard function, $\chi_L(q,\rho_0)$.

Stimulated by the advantages of the
orbital-free {\em ab-initio} simulations,  
there has been a renewed interest in the development of
accurate kinetic energy functionals. 
With Perrot's work as the basis, \cite{Perrot} Madden and coworkers 
\cite{MaddenLRT,MaddenQRT} have developped 
functionals which correctly recover the Thomas-Fermi and linear response
limits \cite{MaddenLRT} and have also included the quadratic 
response. \cite{MaddenQRT}  Later, Carter and coworkers \cite{Carter} 
investigated these functionals and proposed a linear combination of them 
as a suitable form for $T_s$; more recently they have also derived another 
expression which includes density dependent kernels.
Unfortunately, an undesirable feature  
of these functionals is that they are not positive definite, so that 
minimization of the energy functional can lead to an 
unphysical negative kinetic energy.

Chac\'on, Alvarellos and Tarazona
\cite{CAT}  have developed a different type of kinetic energy functional, 
which employs an ``averaged  
density" and recovers the uniform and LRT limits. Their functional has 
been investigated and generalized by 
Garc\'\i a-Gonz\'alez {\em et al}.\cite{PGGs} These 
functionals have the merit of 
being positive definite, but they are somewhat complicated to apply and 
require order $N$ more Fast Fourier Transforms (FFT's) than simpler 
functionals, and this diminishes the advantage of the orbital free approach 
over the full Kohn-Sham method. 

In this paper we use a simplification of the averaged density 
approach, \cite{PGGs} with the kinetic energy given by

\begin{eqnarray}
T_s=T_W [\rho ]+T_{\beta} [\rho]\\
T_{\beta}[\rho] = \frac{3}{10} \int d\vec{r} \, \rho(\vec{r})^{5/3-2\beta}
\tilde{k}(\vec{r})^2 \\
\tilde{k}(\vec{r}) = (2k_F^0)^3 \int d\vec{s} \, k(\vec{s})
w_{\beta}(2k_F^0|\vec{r}-\vec{s}|) \\
k(\vec{r})=(3\pi^2)^{1/3} \rho(\vec{r})^{\beta} \, .
\end{eqnarray}

\noindent where $k_F^0$ is the Fermi wavevector corresponding to a 
mean electron density $\rho_0$, and 
$w_{\beta}(x)$ is a weighting function, determined by requiring the correct 
recovery of the LRT and uniform density limits. 
Note that $\tilde{k}(\vec{r})$ appears as a convolution which can be
performed rapidly by the usual FFT techniques.
This functional is a generalization of one with $\beta = 1/3$, used earlier 
by us in a study of expanded liquid Cs. \cite{sara}

The details of the functional are given in appendix \ref{tsfunc}, 
and its main characteristics
are as follows: (i) $\beta$ is a real positive number whose maximum value 
leading to a 
mathematically well behaved weight function is $\approx$ 0.6, (ii) the 
functional recovers the uniform and LRT limits, and  
is positive definite,   
(iii) when $k_F^0 \rightarrow 0$ because the 
mean electron density vanishes, e.g. for a finite system,
the von Weizs\"acker term is recovered if $\beta=4/9$, whereas for other
values of $\beta$, the limit is $T_W+C T_{\rm TF}$, (iv) for values of
$\beta > 0.5$ it is expected that $\mu(\vec{r})\psi(\vec{r})$,  
which is the driving force for the dynamic minimization of the 
energy, (see eqn. (\ref{driv_force}) ) remains finite even for very small 
electronic densities $\rho(\vec{r})$.

Those two last properties will be important in the case of expanded liquid 
metals because of the appareance of large 
inhomogeneities in the atomic distribution, and therefore in the 
electron density, with regions where it becomes very small. Indeed, this 
situation has already been observed in the {\em ab-initio} simulations 
of expanded liquid Na. \cite{bickham} In systems for which 
the appearance of isolated atoms or clusters is likely the  
von Weizs\"acker term would be appropiate,  
and a functional with a value of $\beta$ as close as possible to 4/9 
would be recommended.

In the present simulations we have used  
$\beta=0.51$, which 
in the limit $\rho_0 \rightarrow 0$ gives C= $0.046$ and guarantees, 
at least for the thermodynamic states considered, 
that $\mu(\vec{r})\psi(\vec{r})$ remains finite 
and not too large everywhere so that the energy minimization 
can be achieved. 

\subsection{Pseudopotentials}
\label{pseudo}

{\it Ab initio} simulations using the full Kohn-Sham 
approach (KS-AIMD) usually
employ nonlocal pseudopotentials obtained by fitting to properties of the 
free atom. \cite{Martins} In an orbital-free approach 
where the electronic density is the  
variable, such non-local pseudopotentials, which  
act differently on different angular momentum components of the 
orbitals, cannot be used.  Instead, local pseudopotentials must be developed 
which include  
an accurate description of the electronic structure in the physical 
circumstances of interest.

When constructing a pseudopotential to be used for a liquid metal, it 
seems more appropriate to use a reference state which closely resembles 
the environment of an atom in the metal, which is quite different from
free space. The pseudopotential used in this simulation has been obtained 
using the neutral pseudoatom (NPA) method \cite{PRE-NPA} in which the 
reference state is an atom at the 
centre of a spherical cavity in the positive background of a uniform 
electron gas. The density of the gas is taken to be the mean 
valence electron density of the system of interest, in our 
case, the liquid metal in a specific thermodynamic state. The radius of the 
cavity is such that the total positive charge removed from the hole is equal 
to the valence of the atom. First, a full Kohn-Sham density functional   
(KS-DFT) calculation is performed 
to obtain the displaced valence electron density, 
$n_{\rm ps}(r)$, i.e. the change in the 
electron density induced by the atom and the cavity. After pseudizing the 
$n_{\rm ps}(r)$ by eliminating the core-orthogonality oscillations,  
an effective local pseudopotential is constructed  
which, when inserted into the uniform electron gas along with 
the cavity, reproduces the  displaced valence electron density 
previously obtained.

Two approaches will be followed in this paper. The first uses 
LRT to reproduce the displaced valence electron density  
leading to a LRT-based local pseudopotential (LRT-PS) from which an 
effective interatomic pair potential is derived (see eqn.(\ref{phind})) to be 
used in CMD simulations; for further details 
we refer to Ref. \onlinecite{PRE-NPA}. The second approach uses 
the orbital-free 
density funtional theory (OF-DFT) to reproduce the displaced 
electron density and it is 
suited for OF-AIMD simulations. 
The development proceeds as follows. 
When the functional derivatives of the energy functional are performed, the 
Euler equation, eqn. (\ref{eulereq}),  for our 
pseudopotential in the jellium-vacancy system becomes 

\begin{equation}
\mu_s(r) + V_{\rm ext}(r) + V_H(r) + V_{\rm xc}(r) - \mu = 0 \, ,
\label{mueq}
\end{equation}

\noindent where each of the terms is the derivative of the corresponding
term in eqn. (\ref{etotal}), namely,

\begin{equation}
\label{mus}
\mu_s(r) = \mu_W(r) + \mu_{\beta}(r)  \, ,
\end{equation}

\noindent with the expressions for the von-Weizs\"acker term and the
$\beta$-term given in appendix \ref{mu_s},

\begin{equation}
V_{\rm ext}(r) = v_{\rm ps}(r) + v_{\rm cav}(r) + v_{\rm jell}(r) \, ,
\end{equation}

\begin{equation}
\label{VHartree}
V_H(r)=\int d\vec{s} \rho(s)/|\vec{r}-\vec{s}| \, ,
\end{equation}

\noindent with $\rho(r)=\rho_0+n(r)$, 
and $V_{\rm xc}(r)$ is the exchange-correlation potential, obtained
from the functional derivative of $E_{\rm xc}[\rho]$ evaluated at $\rho(r)$.

Due to the spherical symmetry of the system all 
the magnitudes depend just on $r$. 
Given $\rho(r)$, $v_{\rm ps}(r)$ can be obtained from 
eqn. (\ref{mueq}), and the constant $\mu$
is just an energy origin which is set so as to obtain a pseudopotential that
decays to zero for large distances.  The pseudopotential 
constructed in this way, will be refered to as the OF-DFT-based 
pseudopotential (OFDFT-PS).

\section{Results and discussion.}
\label{results}

We have performed OF-AIMD simulations for liquid Al 
at two different thermodynamic states 
along the liquid-vapor coexistence line ( 943 K and 1323 K),   
for which X-ray and neutron diffraction data  
are available. \cite{Jovic,Takeda,Waseda,SQM-LIQ} 
Table \ref{states} gives further 
details on the thermodynamic 
states and several simulation parameters. 
In addition, we have also carried out classical MD simulations, using 
effective interionic pair potentials derived from standard 
second order pseudopotential perturbation theory, with the 
LRT-PS's  constructed as previously 
described (see also  Ref. \onlinecite{PRE-NPA}) and with pair potentials 
derived from the QHNC.

In the OF-AIMD simulations 500 particles were treated in a cubic 
cell of the size appropriate to the density, whereas more particles 
were used for the CMD simulations (see Table \ref{states}). 
In both sets of simulations,  
liquid static properties were evaluated (pair distribution
functions and static structure factors) and several dynamic
properties, both single-particle ones (velocity autocorrelation function,
mean square displacement) and collective ones (intermediate scattering
functions, dynamic structure factors, longitudinal and transverse 
currents). The calculation of the collective
dynamic properties required long simulation runs in order to accumulate 
reasonable statistics; for example the OF-AIMD simulations  
lasted for 2 $\cdot 10^4$ steps which correspond to 30 ps 
of simulation time. On the other hand, the CMD simulations run 
for $10^5$ steps, amounting to 200 ps.

\subsection{Pseudopotentials.}

\begin{figure}
\begin{center}
\mbox{\epsfig{file=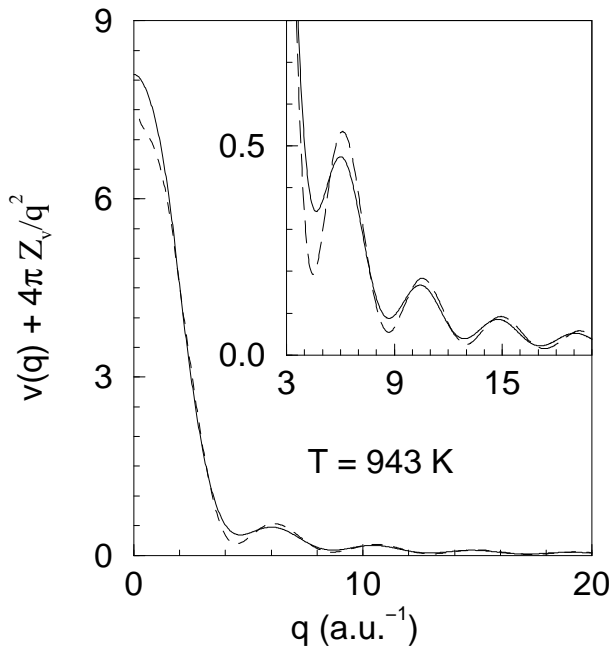,angle=-90,width=80mm}}
\end{center}
\caption{Non-coulombic part of the pseudopotential for Al at T = 943 K. 
The continuous line is the OFDFT-PS  
used in the OF-AIMD simulations, while the dashed line is stands for 
the LRT-PS used for the CMD simulations. }
\label{pseudofig}
\end{figure}

The local pseudopotentials described in section \ref{pseudo} were 
constructed using 
a reference system mimicking the complex system to be
studied. The pseudopotentials  change with the  
thermodynamic state considered and therefore 
are not transferable to other states.  
Figure \ref{pseudofig} shows the Fourier transforms of 
the non-coulombic part of the pseudopotentials obtained from the 
LRT and OF-DFT approaches outlined above. 
The two schemes lead to similar pseudopotentials with the main 
differences being at low $q$-values and in the amplitude of the 
oscillations at large $q$.  Note that in both approaches   
the same pseudized displaced valence electronic 
density of an atom in a jellium-vacancy model is reproduced, although 
the OF-DFT was used in one case and the LRT in the other. 
Consequently the  differences in the two pseudopotentials 
should reflect the importance of
nonlinear effects which, according to the present results seem to 
be more important at small $q$. The appearance of the 
oscillations can be traced back to the calculation of the pseudized 
dispaced valence electronic density which has a discontinuous 
second derivative at a matching radius. However, these oscillations do not 
influence the final OF-AIMD results because they appear for $q$-values 
bigger than those corresponding to the $E_{\rm Cut}$.   
 
%$$$$$$$$$$$$$$$$$$$$$$$$$$$$$$$$$$$$$$$$$$
%$$$$$$$$$$$$$$$$$$$$$$$$$$$$$$$$$$$$$$$$$$

\subsection{Static properties.}

\begin{figure}
\begin{center}
\mbox{\psfig{file=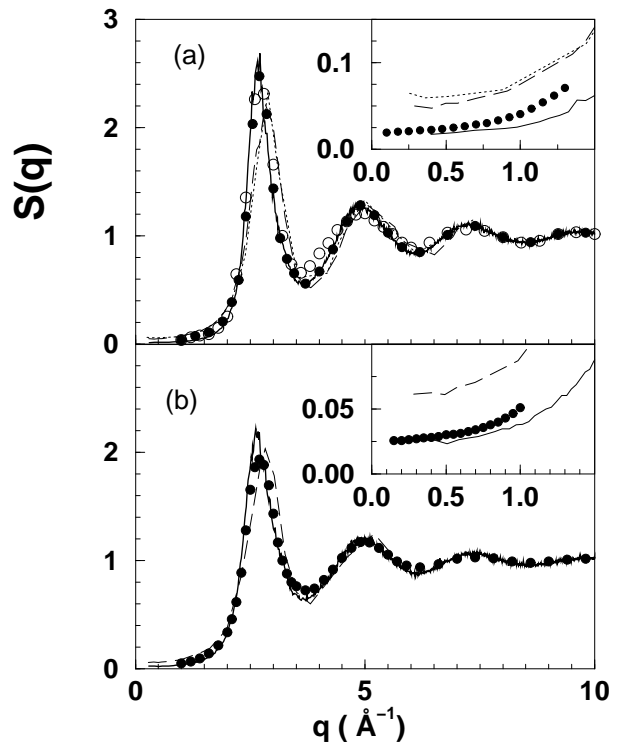,angle=-90,width=80mm}}
\end{center}
\caption{Static structure factors of liquid Al at (a) 943 K and (b) 1323 K. 
Full circles: experimental X-ray diffraction 
data. \protect\cite{Waseda,SQM-LIQ}
Open circles: experimental neutron diffraction 
data. \protect\cite{Takeda}  Continuous line: OF-AIMD simulations. 
Dashed lines: LRT-CMD simulations. Dotted lines: CMD simulations with the QHNC 
potential. The insets show the low-$q$ behaviour.} 
\label{sqfig}
\end{figure}

The static structure factors, $S(q)$, obtained from the simulations
are shown in Figure \ref{sqfig}, which also shows 
the corresponding experimental data measured by  
neutron \cite{Takeda} and X-ray 
diffraction \cite{Waseda,SQM-LIQ} experiments. 
The experimental data show small differences 
in the region $2$ \AA$^{-1}$ $\leq q \leq 5$ \AA$^{-1}$, 
with the neutron values being slightly bigger that the X-ray ones, 
whereas the 
OF-AIMD results stand remarkably well between 
both sets, although somewhat closer to the X-ray data. 
The insets of the figures show that the OF-AIMD results 
in the small $q$-region are also in good agreement with the 
experimental X-ray results. The figures also 
include the $S(q)$, obtained from the CMD simulations performed 
with the interatomic pair potential derived from the LRT-PS and from the 
QHNC method. \cite{Chihara&Kambayasi,Anta&Louis} Although these calculated 
$S(q)$ reasonably reproduce the experimental 
data the agreement with experiment is much better for OF-AIMD.  

%$$$$$$$$$$$$$$$$$$$$$$$$$$$$$$$$$$$$$$$$$$
%$$$$$$$$$$$$$$$$$$$$$$$$$$$$$$$$$$$$$$$$$$

Extrapolation of $S(q)$  to $q \to 0$ allows the isothermal 
compresibility, $\kappa_T$, to be estimated from the 
relation $S(0) = \rho k_B T \kappa_T$. 
A least squares fit of 
$S(q)=s_0+s_2 q^2$ to the calculated $S(q)$ for $q$-values  
up to 1 \AA$^{-1}$ yields the result  
$\kappa_{T, \rm OF-AIMD}$ = 2.37 (in 10$^{-11}$ m$^2$ Nw$^{-1}$ units)  
for T = 943 K, which is close to the experimental value \cite{Seemann} 
$\kappa_{T}$ = 2.43.  In contrast, 
both the LRT-PS and QHNC interionic pair potentials lead 
to much higher values, namely $\kappa_{T, \rm LRT-CMD}$ = 6.5 and  
$\kappa_{T, \rm QHNC-CMD}$ = 7.4  respectively.

The ionic and electronic static structure of liquid 
Al near melting has also been calculated by Anta {\it et al.} \cite{Anta}  
using the OF-AIMD method with a kinetic energy functional   
which describes the correct linear and 
quadratic response of the electron gas \cite{MaddenQRT} and a  
local ionic pseudopotential constructed from a non-local 
ionic pseudopotential. \cite{Anta,Watson} Their results for the 
static structure factor closely followed the experimental one.

\subsection{Dynamic properties.}

\subsubsection{Single-particle dynamics.}

\begin{figure}
\begin{center}
\mbox{\psfig{file=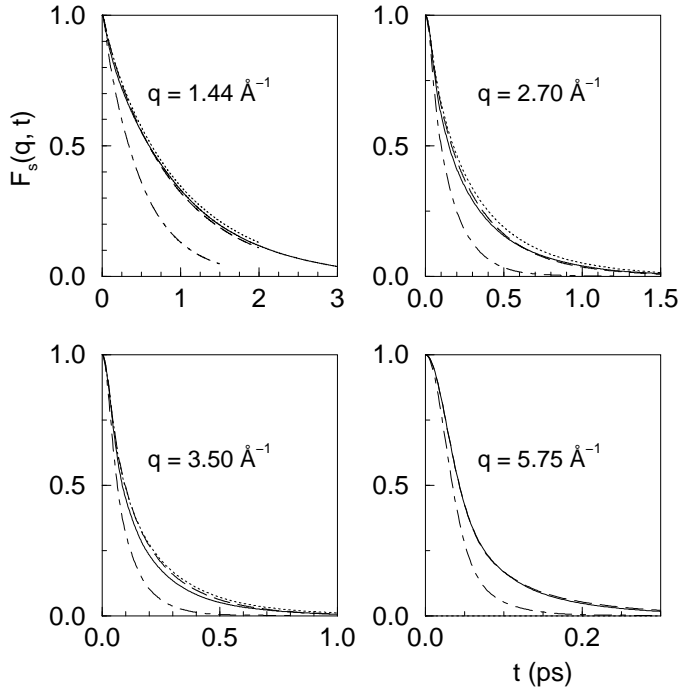,angle=-90,width=90mm}}
\end{center}
\caption{Self intermediate scattering functions, $F_s(q, t)$, at  
several $q$-values, for liquid aluminium.   
Continuous, dashed and dotted lines: OF-AIMD, LRT-CMD and  
QHNC-CMD simulations respectively at T= 943 K. 
Dash-dotted line: OF-AIMD simulations at T=1323 K }
\label{fsqtfig}
\end{figure}

The most complete information about the single-particle properties is 
provided by the self-intermediate 
scattering function, $F_s(q, t)$, which probes 
the single-particle dynamics over different length scales, ranging 
from the hydrodynamic limit ($ q \to 0$) to the free-particle 
limit ($q \to \infty$). In the present simulations, this magnitude has been 
obtained by 

\begin{equation}
F_s(q, t) = \frac{1}{N} \langle \sum_{j=1}^N  
e^{-i {\vec q}{\vec R}_j(t + t_0)} 
  e^{i {\vec q}{\vec R}_j(t_0)} \rangle
\end{equation}

\noindent 
and in figure \ref{fsqtfig} we show the results obtained for several $q$-values 
at T= 943 K and 1343 K. It shows the typical monotonic 
decrease with time; moreover, the results are very similar to 
those of the LRT-CMD and QHNC-CMD simulations, although the latter 
ones show a slightly slower decay with time. An increase in  
temperature leads to increased rate of decay.

%****************************************
%*******************************************************

\begin{figure}
\begin{center}
\mbox{\psfig{file=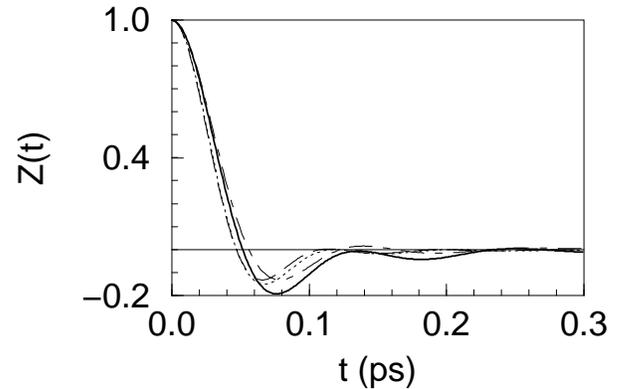,angle=-90,width=80mm}}
\end{center}
\caption{Normalized velocity autocorrelation functions. 
Continuous, dashed and dotted lines: OF-AIMD, LRT-CMD and  
QHNC-CMD simulations respectively at T= 943 K. 
Dash-dotted line: OF-AIMD results 
for T= 1323 K }
\label{fcvfig}
\end{figure}

Closely related to the $F_s(q, t)$ is the 
velocity autocorrelation function (VACF)
of a tagged ion in the fluid, $Z(t)$, which can be 
obtained as the $q \to 0$ limit of the first-order memory function of the 
$F_s(q, t)$. However, in the present simulations it is more easily  
obtained from its definition

\begin{equation}
Z (t) = \langle \vec{v}_1(t) \vec{v}_1(0) \rangle
/ \langle v_1^2 \rangle  
\end{equation}

\noindent which stands for the normalized VACF. The results 
are shown in Figure \ref{fcvfig} along with those derived fron the 
LRT-CMD and QHNC-CMD simulations.  
%$$$$$$$$$$$$$$$$$$$$$$$$$$$$$$$$$$$$$$$$$$
%$$$$$$$$$$$$$$$$$$$$$$$$$$$$$$$$$$$$$$$$$$
The results display the typical backscattering behaviour, which is more 
marked for the OF-AIMD simulations, but the results of the three simulations 
are rather similar. The main 
features of the obtained $Z(t)$ are 
comparable to those obtained for other simple metals near 
melting, \cite{Balubook,Balucani,Li470,Alkterr} namely: (i) a first 
minimum about 0.20 deep and (ii) a rather weak following maximum 
peaking close to cero. The 
self-diffusion coefficient, $D$, is readily obtained from either 
the time integral of $Z(t)$ or from the slope of the mean 
square displacement $\delta R^2(t) \equiv 
\langle | \vec{R}_1(t) - \vec{R}_1(0) |^2 \rangle$  
of a tagged ion in the fluid, as follows

\begin{equation}
D= \frac{1}{\beta m} \int_0^{\infty} Z(t) dt\; ;\hspace{0.91 cm}
D= \lim_{t \to \infty} \delta R^2(t)/6t 
\end{equation}

\noindent
and the results for $D$ are given in Table \ref{diffu}. The two routes 
for $D$ lead to practically the same value, namely 
$D_{\rm OF-AIMD}= 0.49 \;$ \AA$^2$/ps; which is somewhat smaller than 
the mean value of $ 0.55 \;$ \AA$^2$/ps obtained 
in a previous OF-AIMD calculation with 205 particles. \cite{GGLS1}  
Unfortunately, to 
our knowledge, no experimental results are yet available for the 
diffusion coefficients of liquid Al at any thermodynamic state. 
However, we 
can compare with the results of a KS-DFT calculation 
\cite {Blochl} performed for liquid Al near the tripe point, using 64 
particles and a non-local Bachelet-Hamann-Schluter type 
pseudopotential; \cite{Hamann1,Bachelet,Hamann2} this calculation lead to 
a value $D_{\rm KS-DFT}= 0.60 \;$ \AA$^2$/ps derived from the slope of the 
corresponding mean square displacement. 
Recently, another KS-DFT calculation \cite{Alfe&Gillan}
for liquid Al at 1000 K, using 64 particles and ultrasoft 
Vanderbilt pseudopotentials gave for $D$ values within the range 
0.52-0.68 \AA$^2$/ps, derived from the slope of the mean square displacement. 
Our OF-AIMD simulations, 
with a small number of particles and/or a small number of configurations, 
suggest that the self-diffusion coefficients obtained from 
the $\delta R^2(t)$ tend to 
be greater than those obtained by integration of the $Z(t)$, and 
as the number of particles and/or configurations is increased, the  
value for the self-diffusion coefficient is decreased. 
More extensive KS-DFT simulations would probably lead to 
a smaller value of $D$ closer to that obtained in the present 
OF-AIMD simulations.  The values obtained from  
LRT-CMD ($0.58$ \AA$^2$/ps ) and QHNC-CMD ($0.55$ \AA$^2$/ps ) 
simulations are also rather 
similar and slightly bigger than the OF-AIMD result. The 
CMD simulations of Ebbsjo {\it et al.} \cite{Ebbsjo} using several pair 
potentials gave values within the range 0.41-0.45 \AA$^2$/ps. 

\medskip

\begin{figure}
\begin{center}
\mbox{\psfig{file=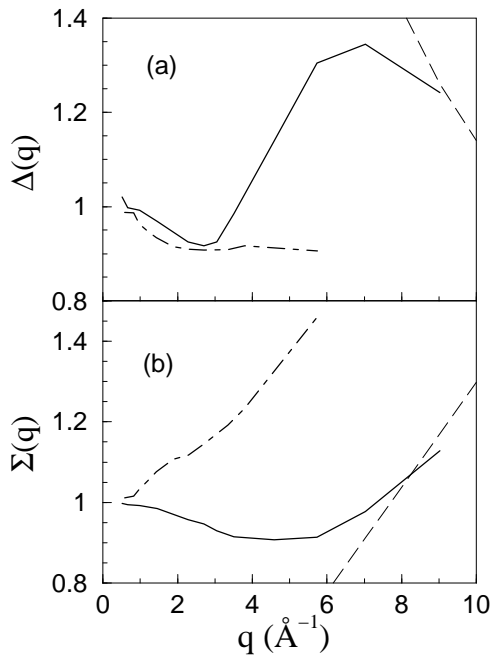,angle=-90,width=65mm}}
\end{center}
\caption{(a) Normalized half-width of $S_s(q, \omega)$, relative to 
its value at the hydrodynamic limit, for liquid aluminium at T = 943 K 
(continuous line) and 1323 K (dash-dotted line). 
The dashed line stands for the free-particle 
limit.  (b) Same as before, but for 
the normalized peak value $S_s(q, \omega=0)$, relative to 
its value at the hydrodynamic limit }
\label{DelSig}
\end{figure}

By Fourier transforming the $F_s(q, t)$ we obtain the 
self-dynamic structure factor, $S_s(q, \omega)$, which for all $q$-values, 
exhibits a monotonic decay with frequency, from a peak value at 
$\omega = 0$.  
$S_s(q, \omega)$ can be characterized by 
the peak value $S_s(q, \omega = 0)$, and the half-width at 
the half-maximum, $\omega_{1/2}(q)$. These parameters are usually reported  
normalized with respect to the values of the hydrodynamic  
($q$ $\to$ 0 ) limit, by introducing the dimensionless quantities 
$\Sigma(q)=\pi  q^2 D S_s(q, \omega =0)$ and 
$\Delta(q) = \omega_{1/2}(q)/ q^2 D$, where  
$\omega_{1/2}(q)/ q^2$ can be interpreted as an
effective $q$-dependent diffusion coefficient $D(q)$. 
For a liquid near the triple point, $\Delta(q)$ usually exhibits an 
oscillatory behaviour whereas, in a dense gas it decreases monotonically 
from unity at $q$ = 0 to the $1/q$ behaviour at large $q$.  
Fig. \ref{DelSig} shows the OF-AIMD results for  
$\Delta(q)$ and $\Sigma(q)$; the corresponding results from the 
LRT-CMD and QHNC-CMD simulations are rather similar and 
are not shown. 
%***********************************************
%%**************************************************************
The results for $\Delta_{\rm OF-AIMD}(q)$, show that for both  
temperatures, the hydrodynamic limit
is reached from below, with a minimum at around $q \approx q_p$, followed 
by a maximum and by a gradual transition, for greater $q$-values, 
to the free-particle limit. Notice that for the higher temperature, the 
oscillations are heavily damped and the free particle limit is 
approached quickly.   
This oscillating behaviour of $\Delta_{\rm OF-AIMD}(q)$ for
small and intermediate $q$-values has been reported by several
authors and has been attributed to the coupling of the single-particle motion
to other modes in the 
system. \cite{Balubook,MorGla,WahnSjo,Verk1,Verk2,Montfrooy} 
On the other hand, the results for $\Sigma(k)$ reflect greater sensitivity to 
changes in temperature, with the diffusive limit reached from below for 
T=943 K and from above for T=1323 K. 
We note that similar features to those obtained in this paper
were obtained earlier by
Torcini {\it et al.} \cite{TorBalVer} in their CMD study of liquid 
lithium near melting using the interatomic pair potential proposed by    
Price {\it et al}. \cite{PST}

\subsubsection{Collective dynamics.}

\begin{figure}
\begin{center}
\mbox{\psfig{file=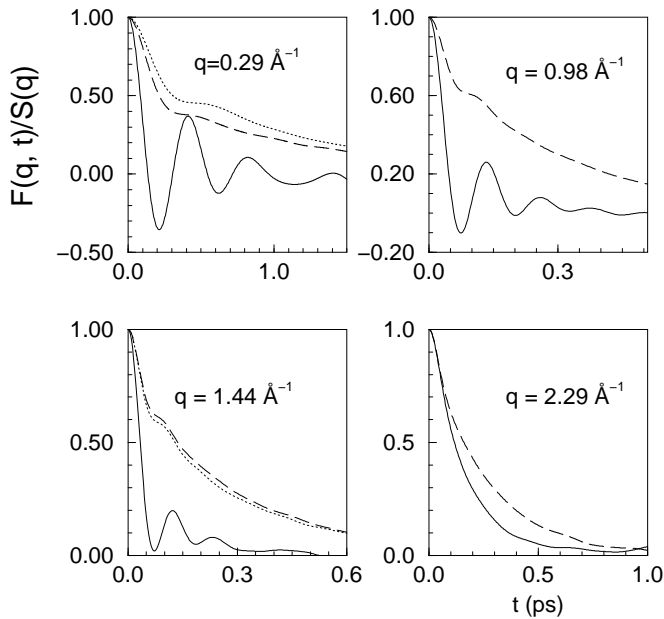,angle=-90,width=87.5mm}}
\end{center}
\caption{Normalized intermediate scattering functions, $F(q, t)$, at  
several $q$-values, for liquid 
aluminium at T = 943 K. 
Continuous line: OF-AIMD simulations.  
Dashed line: LRT-CMD results. Dotted line: QHNC-CMD results.  }
\label{fqtfig1}
\end{figure}

\begin{figure}
\begin{center}
\mbox{\psfig{file=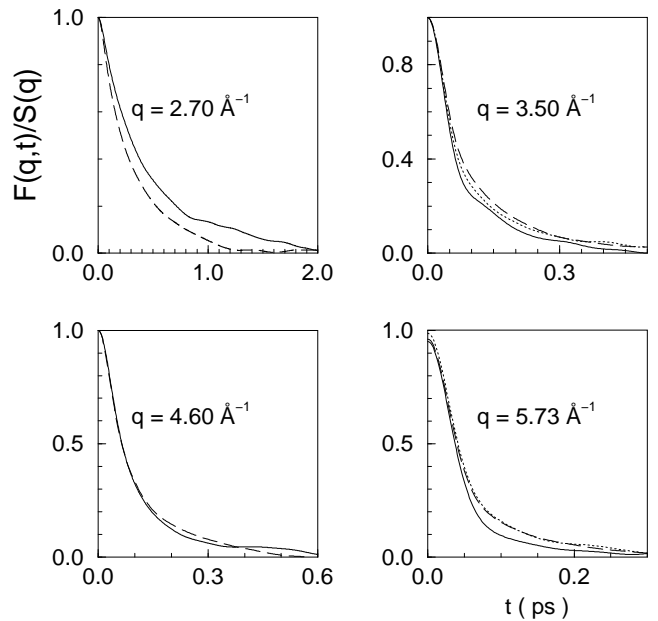,angle=-90,width=85mm}}
\end{center}
\caption{Same as the previous figure} 
\label{fqtfig2}
\end{figure}

The intermediate scattering function, $F(q, t)$, embodies the information 
concerning the collective dynamics of density fluctuations over both the 
length and time scales. It is defined as

\begin{equation}
F(q, t) = \frac{1}{N} \left \langle \left( \sum_{j=1}^N  
e^{-i {\vec q}{\vec R}_j(t + t_0)} \right)
 \; \left( \sum_{l=1}^N e^{i {\vec q}{\vec R}_l(t_0)} \right) \right \rangle
\end{equation}

\noindent and in figures \ref{fqtfig1}-\ref{fqtfig2} we show, 
the results from the present OF-AIMD simulations
for several $q$-values.  
$F(q, t)$ exhibits oscillatory behaviour which persists 
up to $ q \approx 3 q_p/5$, with the amplitude of the oscillations 
being stronger for the smaller $q$-values. This is typical 
behaviour found for other simple liquid metals near melting, 
by either computer simulations \cite{TorBalVer,Shimojo2,Kambayashi} or 
theory. \cite{Casas2} Different behaviour is seen for the results  
in the same $q$-range obtained from the LRT-CMD and QHNC-CMD simulations, 
with $F(q, t)$'s whose contact values, given by $F(q, t=0) = S(q)$, 
are more than double   
and, more important, with a diffusive component playing a 
dominant role.    
The corresponding MD results of 
Ebbjso {\it et al.} \cite{Ebbsjo} for the $F(q, t)$'s  
have better contact values but also display an 
important diffusive component.

%**************************************************
%************************************************************

%**************************************************
%************************************************************

\begin{figure}
\begin{center}
\mbox{\psfig{file=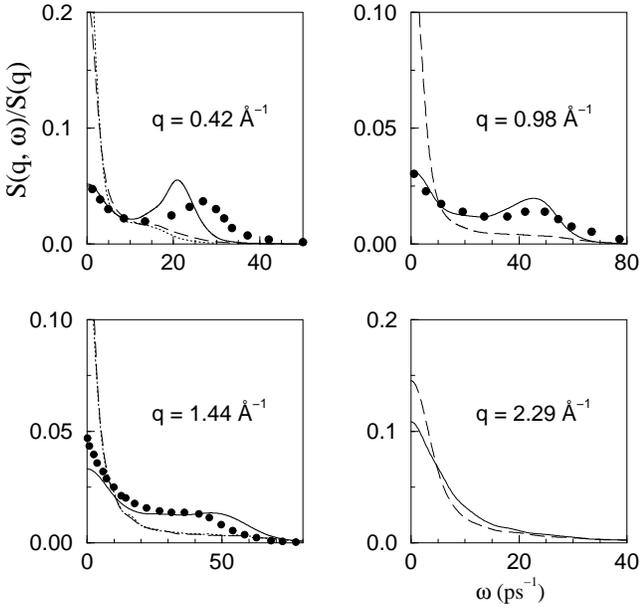,angle=-90,width=85mm}}
\end{center}
\caption{Dynamic structure factor, $S(q, \omega)$, for several
$q$-values, for liquid 
aluminium at T = 943 K. 
Continuous line: OF-AIMD simulations.  
Dashed line: LRT-CMD results. Dotted line: QHNC-CMD results. 
Full circles: Experiment \protect\cite{Scopigno}.  }
\label{sqwfig1}
\end{figure}

\begin{figure}
\begin{center}
\mbox{\psfig{file=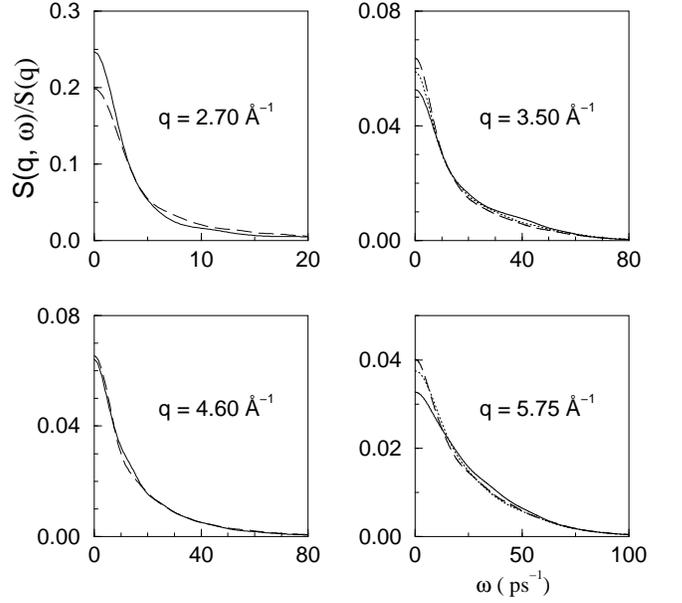,angle=-90,width=85mm}}
\end{center}
\caption{Same as the previous figure} 
\label{sqwfig2}
\end{figure}

Closely connected to the $F(q, t)$ is the dynamic 
structure factor, $S(q, \omega)$, which is obtained by  
a time Fourier transform of the $F(q, t)$ (with an
appropriate window to smooth out truncation effects). Its importance 
lies in its direct connection to the 
inelastic neutron scattering or the IXS data. 
The results obtained for 
the $S(q, \omega)$ are shown in 
figures  \ref{sqwfig1}-\ref{sqwfig2} for a range of 
wavevectors up to $\approx 2.5 q_p$. 
The dynamic structure factor shows well defined sidepeaks, indicative 
of collective density excitations, up to 
$q \approx 1.6$ \AA$^{-1}$ which amounts to $\approx 3 q_p/5$. 
The results qualitatively reproduce the  
shape of the experimental IXS data  \cite{Scopigno}  with some 
small discrepancies in the heigths and positions of the peaks. 
Similar results, but with a better description of the central 
peak at the lowest $q$-values, were also obtained in the CMD 
simulations of Ebbsjo {\it et al}. \cite{Ebbsjo} However, it must 
be stressed that their $F(q, t)$'s  
were previously fitted to an analytical expression interpolating among  
the ideal gas, viscoelastic and hydrodynamic models, and therefrom 
the corresponding $S(q, \omega)$ were derived. 
The strongly diffusive character of the $F(q, t)$'s obtained from 
both the LRT-CMD and QHNC-CMD simulations, give rise to  
$S(q, \omega)$ which decay rather quickly, with hardly discernable 
side peaks. This is because the side peaks are located at smaller  
positions, given by $q c_s(q)$,   
where $c_s(q)$ is the generalized adiabatic sound velocity (see below), 
which is too small because of the large values of 
$S(q)$ at those $q$-values.

\begin{figure}
\begin{center}
\mbox{\psfig{file=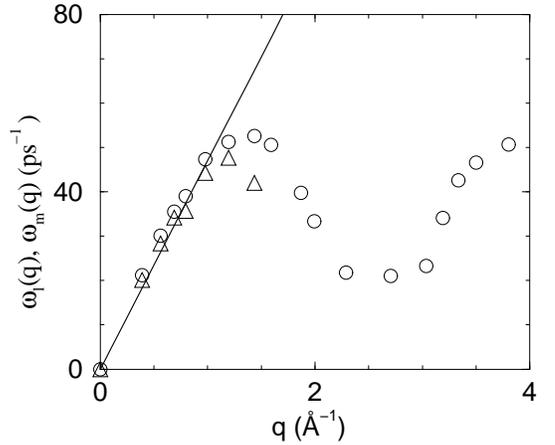,angle=-90,width=70mm}}
\end{center}
\caption{Dispersion relation for liquid Al at T =943 K. Open triangles: 
peak positions, $\omega_m(q)$, from the OF-AIMD $S(q, \omega)$. 
Open circles: peak positions, $\omega_l(q)$, from the maxima of the  
OF-AIMD longitudinal current, $J_l(q, \omega)$. Full line: 
Linear dispersion with the 
hydrodynamic sound velocity, v=4700 m/s. }
\label{disperfig}
\end{figure}

From the positions of the sidepeaks, $\omega_m(q)$, the dispersion relation 
of the density fluctuations has been obtained and this is shown 
in figure \ref{disperfig}  for the state at T = 943 K, 
along with $\omega_l(q)$, which is 
the dispersion relation obtained from the maxima of the longitudinal 
current correlation function, $J_l(q, \omega) = \omega^2 S(q, \omega)$. 
Note that in the hydrodynamic region (small $q$), the slope of the 
dispersion relation curve is the adiabatic sound velocity,  
$c_s(q) = v_{th} \sqrt {\gamma/S(q)}$, with $v_{th}=(\beta m)^{-1/2}$ 
being the thermal 
velocity and $\gamma$ is the ratio of the specific heats. In 
the limit $q \to 0$, $c_s(q)$ reduces to the bulk adiabatic sound velocity 
and determines the slope of the dispersion at $q \to 0$. 
By extrapolating the OF-AIMD results for $S(q)$
and using the experimental value  \cite{Webber-Stephens} of  
$\gamma$ $\approx 1.25$,   
we obtain a value of $\approx 4850$ m/s for the bulk adiabatic 
sound velocity which compares reasonably well with the experimental  
value \cite{Seemann} of $\approx 4700$ m/s, near the triple point. 
Figure \ref{disperfig} shows a {\it positive dispersion}, i.e. an increase 
of $\omega_l(q)$ with respect to 
the values predicted by the hydrodynamic adiabatic speed of sound, with 
a maximum located around 0.4 \AA$^{-1}$. Similar behaviour has also 
been obtained by 
Scopigno {\it et al.} \cite {Scopigno} from their 
experimental IXS results for liquid Al at T=1000 K  
and has been observed 
in other liquid metals: Rb, Cs, 
Li and Na. \cite{Scopigno,BurkelSinn,Pilg0}

Another interesting dynamical magnitude is the transverse  
current time correlation function, $J_t(q,t)$, which is not associated with 
any measurable quantity and can only be determined by means of MD 
simulations. It provides information on the shear modes and is defined 
as

%$$$$$$$$$$$$$$$$$$$$$$$$$$$$$$$$$$$$$$$$$$
\begin{figure}
\begin{center}
\mbox{\psfig{file=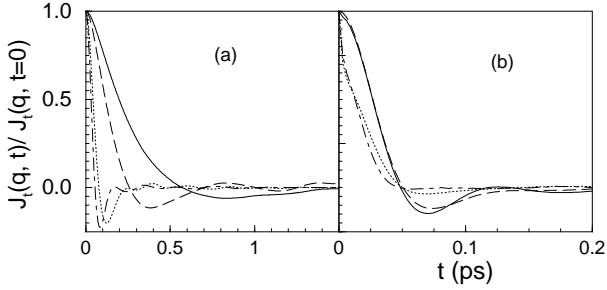,angle=-90,width=80mm}}
\end{center}
\caption{OF-AIMF transverse current correlation function, $J_t(q, t)$, 
at several $q$-values (in \AA$^{-1}$) for liquid Al at T =943 K. 
(a) $q$=0.29 (full curve), 
$q$=0.42 (dashed curve), $q$=0.98 (dotted curve), 
$q$= 1.45 (dash-dotted curve). (b) $q$=2.70 (full curve), 
$q$= 3.80 (dashed curve), $q$= 5.54 (dotted curve), 
$q$= 9.02 (dot-dashed curve). } 
\label{ctqtfig}
\end{figure}

\begin{figure}
\begin{center}
\mbox{\psfig{file=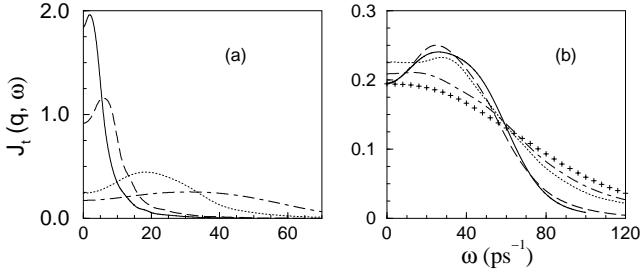,angle=-90,width=85mm}}
\end{center}
\caption{OF-AIMF transverse current correlation spectrum, $J_t(q, \omega)$, 
at several $q$-values (in \AA$^{-1}$) for liquid Al at T =943 K. 
(a) $q$=0.29 (full curve), 
$q$=0.42 (dashed curve), $q$=0.98 (dotted curve), 
$q$= 1.99 (dash-dotted curve). (b) $q$=2.70 (full curve), 
$q$= 3.80 (dashed curve), $q$= 5.54 (dotted curve), 
$q$= 7.03 (dot-dashed curve), $q$= 9.02 (pluses).}  
\label{ctqwfig}
\end{figure}

%$$$$$$$$$$$$$$$$$$$$$$$$$$$$$$$$$$$$$$$$$$

\begin{equation}
J_t(q, t) = \frac{1}{N} \left \langle j_x^*(q,0) j_x(q,t)   
\right \rangle
\end{equation}

\noindent 
where
$j_x(q,t) = \sum_{j=1}^N v_j^x(t) e^{-i {\vec q}{\vec R}_j(t)}$  
is the transverse current. The shape of $J_t(q, t)$ evolves from a 
gaussian, in both $q$ and $t$, for the free particle $q \to \infty$ limit,  
towards a gaussian in $q$ and  exponential in $t$ for the hydrodynamic 
limit ($q \to 0$), i.e. 

\begin{equation}
J_t(q \to 0, t) = \frac{1}{\beta m} e^{-q^2 \eta \mid t \mid /m \rho}
\label{Jtqthyd}
\end{equation}

\noindent 
where $\eta$ is the shear viscosity. For intermediate $q$-values, 
$J_t(q, t)$ exhibits a more complicated behaviour, as shown in 
Figure \ref{ctqtfig} where OF-AIMD results for 
liquid Al near melting are shown. Note that for the smallest $q$-value
reached by the simulation:  $q$ = 0.29 \AA$^{-1}$, the 
corresponding $J_t(q, t)$ takes on negative values, which by 
eqn. (\ref{Jtqthyd}) means that it is 
already  beyond the hydrodynamic regime. 
The associatted spectrum, $J_t(q, \omega)$,  plotted in 
Figure \ref{ctqwfig}, shows an inelastic peak which already exists  
at $q$ = 0.29 \AA$^{-1}$ $\approx$ $0.11 q_p$ ;  as $q$ increases the peak  
becomes better defined and it persists to $q$ values 
around $3 q_p$, although it 
has already disappeared for the largest $q$-value considered. 
Note that the associatted 
peak frequency increases with $q$ up a maximum value at 
$q \approx q_p$, and then flattens at larger $q$ as $J_t(q, \omega)$ 
evolves towards a gaussian shape. This behaviour closely parallels that
observed for the alkali metals where the inelastic peak appears 
for $q \geq 0.07 q_p$. \cite{Balubook} 

Similar results are also obtained by the LRT-CMD approach,  
but $J_t(q, t)$ decays more slowly and the minima are less 
marked. This leads to a spectrum $J_t(q, \omega)$ where the peaks 
are less marked and, in fact, there is no peak at 
$q$ = 0.29 \AA$^{-1}$. 

%$$$$$$$$$$$$$$$$$$$$$$$$$$$$$$$$$$$$$$$$$$
%$$$$$$$$$$$$$$$$$$$$$$$$$$$$$$$$$$$$$$$$$$

%$$$$$$$$$$$$$$$$$$$$$$$$$$$$$$$$$$$$$$$$$$
%$$$$$$$$$$$$$$$$$$$$$$$$$$$$$$$$$$$$$$$$$$

%$$$$$$$$$$$$$$$$$$$$$$$$$$$$$$$$$$$$$$$$$$
%$$$$$$$$$$$$$$$$$$$$$$$$$$$$$$$$$$$$$$$$$$

%$$$$$$$$$$$$$$$$$$$$$$$$$$$$$$$$$$$$$$$$$$
%$$$$$$$$$$$$$$$$$$$$$$$$$$$$$$$$$$$$$$$$$$

From the results  for $J_t(q, t)$ we can readily 
obtain the shear viscosity coefficient, $\eta$ as 
follows. \cite{Balubook,Palmer,BaBroJedVa}  
The memory function representation of the $J_t(q,  t)$:  

\begin{equation}
\tilde{J_t}(q, z)= \frac{1}{\beta m} 
\left [ z + \frac{q^2}{\rho m} \;  \tilde{\eta}(q, z)\right ]^{-1}
\end{equation}

\noindent where the tilde denotes the Laplace transform, introduces a  
generalized shear viscosity coefficient, $\tilde{\eta}(q, z)$. 
The area under the normalized 
$J_t(q, t)$, gives  
$\beta m \; \tilde{J_t}(q, z=0)$ from which values 
for $\tilde{\eta}(q, z=0)$ can be obtained which when 
extrapolated to $q=0$ give  
the usual shear viscosity coefficient, $\eta$. Results for 
$\eta$ presented in Table \ref{diffu} 
compare favourably with the available 
experimental data. \cite{Shimoji&Itami} For comparison, we note that the 
KS-DFT simulations of Alfe and Gillan \cite{Alfe&Gillan} gave values 
in the range 1.4 - 2.2 GPa $\cdot$ ps.

\section{Conclusions.}

Several dynamic properties of liquid aluminium have been calculated  
at two thermodynamic states close to the triple point. The simulations have 
been performed using the orbital free {\em ab-initio} molecular dynamics 
method, showing the feasibility 
of this technique to calculate several 
time correlation functions, allowing a comprehensive study of the 
dynamical properties. Furthermore, agreement with the available 
experimental data is quite satisfactory.

We have also presented a method for producing, from first principles, 
local pseudopotentials for use with the orbital free 
density functionals.
While the ultimate goal of the method would be to use the atomic number 
of the atoms as the only input data, this has not yet been achieved as the 
present calculations also require the experimental number density of the 
system for calculating the local pseudopotential 
and for performing the simulations. However, we stress that starting 
from very basic information, the present scheme allows the determination 
of the static and dynamic properties of the system. 

Finally, we emphasize that in the present scheme, the calculation 
of the pseudopotential is coupled to the particular functional  
adopted for the total potential energy of the system. This means 
that different kinetic energy functionals would lead to different 
pseudopotentials. Consequently, this field is open to further 
improvements in the 
description of the kinetic energy and, therefore, also in the 
corresponding local pseudopotential. 

\bigskip

\section*{Acknowledgements}

This work has been supported by the Junta de Castilla 
y Le\'on (Project No. VA70/99), NATO (CRG971173) and the 
DGES (PB98-0641-C02-01).  DJG acknowledges the 
UVA for the provision of financial support to visit the 
the Physics Dpt. of Queen's University were part of this work was 
carried out. MJS acknowledges the support of the NSERC of Canada.

\appendix

\section{The kinetic energy functional}
\label{tsfunc}

We consider the kinetic energy functional 

\begin{equation} 
T_s[\rho] = T_{W}[\rho] + T_{\beta}[\rho] 
\end{equation}

\noindent where 

\begin{equation} 
T_{\beta}[\rho]= \frac3{10} \int d\vec{r} \, \rho(\vec{r})^{5/3-2\beta}
\tilde{k}(\vec{r})^2 
\label{eq2}
\end{equation}

\begin{equation} 
\tilde{k}(\vec{r}) = (2k_F^0)^3 \int d\vec{s} \, 
k(\vec{s}) \,
\omega_{\beta}(2 k_F^0
|\vec{r}-\vec{s}|) \equiv k(\vec{r})*\omega_{\beta}(2k_F^0r)
\label{eqconv}
\end{equation}

\begin{equation}
 k(\vec{r})= (3\pi^2)^{1/3} \rho(\vec{r})^{\beta} 
\end{equation}

In the limit of small deviations from a uniform system, we wish to 
recover the LRT result. Equating the Fourier 
transform (FT) of the 
second functional derivative of $T_s[\rho]$ with respect to 
$\rho(\vec{r})$   
for $\rho(\vec{r})=\rho_0$, to the inverse of the 
Lindhard response function, gives for the 
weight function

\[ \left(6\beta^2-\frac{20}{3}\beta+\frac{10}{9}\right) 
+ 4 \beta \, \left(\frac53-2\beta\right) \, \overline{\omega}_{\beta}(\eta) + 
2 \beta^2 \, \overline{\omega}_{\beta}(\eta)^2  =  \]
\begin{equation}
 \frac{10}{9}\left(1/\pi_L(\eta) - 3 \eta^2\right)
\label{eq6}
\end{equation}

\noindent where $\eta=q/2k_F^0$, $\overline{\omega}_{\beta}$ is the FT  
of $\omega_{\beta}$  and 

\begin{equation} \pi_L(\eta) = \frac12 \left( 1 +
\frac{1-\eta^2}{2\eta} \ln \left| \frac{1+\eta}{1-\eta} \right| \right) \, .
\end{equation}

\noindent is the noninteracting homogeneous electron gas response function. 
Taking in eqn. (\ref{eq6}) the solution 
which satisfies  
the normalization condition $\overline{\omega}_{\beta}(\eta=0) = 1$, 
and with $\beta$ within the range $0 \le \beta \le 5/6$ so that the 
power of $\rho(\vec{r})$ in eqn. (\ref{eq2}) is positive, 
the weight function is given by

\begin{equation}
\overline{\omega}_{\beta}(\eta)=
2 - \frac{5}{3\beta} + \frac{1}{3\beta}\sqrt{(5-3\beta)^2 +
5 (\pi_L^{-1}(\eta)-1-3\eta^2)} \, .
\end{equation}

Requiring $\overline{\omega}_{\beta}$ to be real 
places a stricter limit on $\beta$: $\beta \leq 0.5991$. 
With this choice of weight function, the  
functional recovers the LRT limit, and in the limit of 
uniform density it reduces to the Thomas-Fermi functional. 
In the limit $\eta\rightarrow\infty$ we have 

\begin{equation}
\overline{\omega}_{\beta}(\eta) \rightarrow C_1 + A/\eta^2 + \cdots 
\end{equation}
where

\begin{equation}
C_1 = 2 - \frac5{3\beta}+\frac1{3\beta}\sqrt{17-30\beta+9\beta^2}
\end{equation}

The constant $C_1$ gives rise to a Dirac delta function in
the real space; therefore it is convenient 
to define a ``modified" weight function

\begin{equation}
 \tilde{\omega}_{\beta}(\eta) = \overline{\omega}_{\beta}(\eta) - C_1 
\end{equation}
so that every convolution involving $\omega_{\beta}$, such as 
in eqn. (\ref{eqconv}), becomes 

\begin{equation}
 G(\vec{r})*\omega_{\beta}(2k_F^0r) = C_1 G(\vec{r}) + 
G(\vec{r}) * \tilde{\omega}_{\beta}(2k_F^0r) \, .
\end{equation}

An important limit is when the mean electron density, and therefore 
$k_F^0$, vanishes as for instance in a finite system. 
Now, the convolutions involving the ``modified" weight function 
vanish because $\eta = q/2k_F^0 \to \infty$ and  
$\tilde{\omega}(\eta)$ vanishes.
Consequently, $\tilde{k}(\vec{r})=C_1 \, k(\vec{r})$, and the kinetic 
energy functional becomes $T_s[\rho]=T_{W}[\rho]+C_1^2 \; T_{TF}[\rho]$, 
and when $\beta=4/9$, $C_1=0$.

\section{The position dependent chemical potential.}
\label{mu_s}

The functional derivative of eqn. (\ref{etotal}) gives 

\begin{equation}
\mu(\vec{r}) = \mu_W(\vec{r}) + 
\mu_{\beta}(\vec{r}) +
V_{\rm ext}(\vec{r}) + 
V_H(\vec{r}) +
V_{xc}[\rho(\vec{r})] 
\end{equation}

\noindent where

\begin{equation}
\mu_W(\vec{r}) = \frac18
\frac{\left|\vec{\nabla} \rho(\vec{r})\right|^2}
{\rho(\vec{r})^2} -
\frac14 \frac{\nabla^2\rho(\vec{r})}{\rho(\vec{r})} \, ,
\end{equation}

\begin{equation}
V_H(\vec{r}) = 
\int d\vec{s} \, \frac{\rho(\vec{s})}{\left|\vec{r}-\vec{s}\right|} \, ,
\end{equation}
and in terms of the modified weight function 

\begin{eqnarray}
\mu_{\beta}(\vec{r}) = \frac3{10} &  
[ \,\, \left(5/3-2\beta\right)\rho(\vec{r})^{2/3-2\beta} 
\tilde{k}(\vec{r})^2+  \nonumber \\
 & 2\beta (3\pi^2)^{1/3} \rho(\vec{r})^{\beta-1} h(\vec{r}) \,\, ]
\end{eqnarray}
with

\begin{equation}
h(\vec{r}) = f(\vec{r}) *  
%[\; \tilde{k}(\vec{r})  \; \rho(\vec{r})^{5/3-2\beta} \; ] *  
\tilde{\omega}_{\beta}(2k_F^0r)
\end{equation}

\noindent where 
$f(\vec{r}) = \; \tilde{k}(\vec{r})  \; \rho(\vec{r})^{5/3-2\beta} \;$.    
The product $\mu(\vec{r})\psi(\vec{r})$, is the ``driving
force" for the dynamical minimization of the energy 
functional, see eqn. (\ref{driv_force}). 
If the various powers of the density appearing in  
$\mu_{\beta}(\vec{r})$$\psi(\vec{r})$ are to remain positive so that 
this driving force is not to diverge in regions where the density vanishes, 
then $1/2 \le \beta \le 7/12$. 
In practice we have found that for $\beta=0.51$ the minimization
has always proved possible.

\section{Constructing the local pseudopotential from an infinite system}
\label{constr_ps}

In an infinite system most  of the previous expressions diverge,
because the integrals extend to  all space, and the integrands 
don't vanish for large distances.
Moreover, the normalization constraint must be redefined, because the total
number of electrons is infinite. 

To avoid these problems, one has to take into account
the similar divergencies which appear in the ``ionic" part of the
total energy. This amounts to  
use the ``difference" functions, which
are obtained by substracting from the total functions their corresponding
limits for large distances, and redefining the chemical potential as the
Lagrange multiplier associated to the normalization of the ``displaced"
density $n(\vec{r})=\rho(\vec{r})-\rho_0$.
In this way we define the following functions:

\begin{itemize}
\item $\Delta_{\alpha}(\vec{r}) = \rho(\vec{r})^{\alpha} - \rho_0^{\alpha} =
(\rho_0 + n(\vec{r}) )^{\alpha}-\rho_0^{\alpha}$

In particular, for $\alpha=1$, $\Delta_1(\vec{r})=n(\vec{r})$, the
displaced density.

\item $\chi(\vec{r}) = k(\vec{r})-
(3\pi^2)^{1/3}\rho_0^{\beta}$

\item $\tilde{\chi}(\vec{r}) = \tilde{k}(\vec{r})-
(3\pi^2)^{1/3}\rho_0^{\beta}$

\item $\phi(\vec{r}) = f(\vec{r}) -  
%[\; \tilde{k}(\vec{r})  \; \rho(\vec{r})^{5/3-2\beta} \; ] -   
(3\pi^2)^{1/3}\rho_0^{5/3-\beta}$

\item $\eta_2(\vec{r}) = h(\vec{r}) -  
%[ f(\vec{r}) * \tilde{\omega}_{\beta}(2k_F^0r) \,] -  
(1-C_1) (3\pi^2)^{1/3}\rho_0^{5/3-\beta}$

\item $\tilde{\mu}_{\beta}(\vec{r}) = \mu_{\beta}(\vec{r}) - \frac12 (k_F^0)^2$

\item $v_{\rm ext}(\vec{r}) = V_{\rm ext}(\vec{r}) - v_{\rm jell}(\vec{r})$, 
where $v_{\rm jell}(\vec{r})$ is the potential created by a uniform 
background of positive charge with density $\rho_0$.

\item $v_{xc}(\vec{r}) = V_{xc}[\rho_0+n(\vec{r})] - V_{xc}[\rho_0]$

\end{itemize}

In terms of these functions,
the Euler equation now becomes: 

\[
v_{\rm ext}(\vec{r})+\int d\vec{s} \, n(\vec{s})/\left|\vec{r}-\vec{s}\right| +
v_{\rm xc}(\vec{r}) + 
\frac18 \frac{\left|\vec{\nabla}n(\vec{r})\right|^2}{(\rho_0+n(\vec{r}))^2}
\]
\begin{equation}
-\frac14 \frac{\nabla^2n(\vec{r})}{\rho_0+n(\vec{r})} +
\tilde{\mu}_{\beta}(\vec{r}) -\mu' = 0
\label{eulereqinf}
\end{equation}
with

\begin{equation}
\tilde{\mu}_{\beta}(\vec{r}) = 
\frac3{10} \left[ \,
\tilde{\mu}_A(\vec{r}) +
\tilde{\mu}_B(\vec{r}) +
\tilde{\mu}_C(\vec{r}) +
\tilde{\mu}_D(\vec{r}) \, \right] \,\,\,\, ,
\label{eqmub}
\end{equation}

\noindent where

\begin{equation}
\tilde{\mu}_A(\vec{r}) =
\left(\frac53-2\beta\right) 
\rho_0^{2/3-2\beta} 
\left[ 2(3\pi^2)^{1/3}\rho_0^{\beta}
\tilde{\chi}(\vec{r}) + 
\tilde{\chi}(\vec{r})^2 \right] \,\,\,\, ,
\end{equation}

\begin{equation}
\tilde{\mu}_B(\vec{r}) =
\left(\frac53-2\beta\right) 
\Delta_{2/3-2\beta}(\vec{r})  
\left[ (3\pi^2)^{1/3}\rho_0^{\beta}+\tilde{\chi}(\vec{r}) \right]^2
\,\,\,\, ,
\end{equation}

\[
\tilde{\mu}_C(\vec{r}) =
2\beta C_1 (3\pi^2)^{1/3} [ \,
\rho_0^{2/3-\beta} \tilde{\chi}(\vec{r}) + 
(3\pi^2)^{1/3}\rho_0^{\beta}\Delta_{2/3-\beta}(\vec{r}) 
\]
\begin{equation}
+ \,
\Delta_{2/3-\beta}(\vec{r}) \tilde{\chi}(\vec{r}) \, ] \,\,\,\, ,
\end{equation}

\noindent and

\[
\tilde{\mu}_D(\vec{r}) =
2\beta(1-C_1)  (3\pi^2)^{2/3} \rho_0^{5/3-2\beta} \Delta_{\beta-1}(\vec{r}) 
\]
\begin{equation}
 + 2\beta (3\pi^2)^{1/3}  \left[\rho_0+n(\vec{r})\right]^{\beta-1} 
\eta_2(\vec{r}) \,\,\,\, .
\label{eqmubfin}
\end{equation}

\noindent In the last equation, $\eta_2(\vec{r})= \phi(\vec{r}) *
\tilde{\omega}_{\beta}(2k_F^0r)$ and 

\[
\phi(\vec{r}) = \rho_0^{5/3-2\beta} \tilde{\chi}(\vec{r}) + 
(3\pi^2)^{1/3} \rho_0^{\beta} \Delta_{5/3-2\beta}(\vec{r}) 
\]
\begin{equation}
+ \Delta_{5/3-2\beta}(r) \tilde{\chi}(\vec{r})
\label{eqfi}
\end{equation}

\bigskip

Summarizing, to evaluate $\tilde{\mu}_{\beta}(\vec{r})$ from the
displaced density $n(\vec{r})$ we take the following steps:

\begin{enumerate}
\item Compute $\Delta_{\alpha}(\vec{r})$ for
$\alpha=5/3-2\beta,2/3-2\beta,2/3-\beta$, and $\beta-1$

\item Compute $\chi(\vec{r})$ and FT to obtain $\chi(\vec{q})$

\item Compute $\tilde{\chi}(\vec{q})=C_1 \chi(\vec{q}) + \chi(\vec{q})
\tilde{\omega}_{\beta}(q/2k_F^0)$ and by inverse 
FT obtain $\tilde{\chi}(\vec{r})$

\item Compute $\phi(\vec{r})$ according to eqn. (\ref{eqfi}) and 
 FT to obtain $\phi(\vec{q})$

\item Compute $\eta_2(\vec{q}) = \phi(\vec{q})
\tilde{\omega}_{\beta}(q/2k_F^0)$ and inverse 
FT to obtain $\eta_2(\vec{r})$

\item Apply eqns. (\ref{eqmub}-\ref{eqmubfin}) to obtain 
$\tilde{\mu}_{\beta}(\vec{r})$

\end{enumerate}

When our system is 1 atom in a jellium-vacancy 
the external potential is given by

\[
v_{\rm ext}(\vec{r})=v_{\rm ps}(r) + v_{\rm cav}(r)
\]
i.e., the sum of the potential created by the cavity and the
ionic pseudopotential. 
Substituting into eqn. (\ref{eulereqinf}) we directly obtain $v_{\rm ps}(r)$
once we compute all the other terms, which are calculated from $n(r)$.

Note that all the functions have spherical symmetry, which leads 
to simple expressions for the gradient, the laplacian and also the 
Fourier tranforms.

\newpage

\begin{table}
\caption{Thermodynamic states studied in this work, along with some 
simulation details.} 
\label{states}
\begin{tabular}{cccccc}
  & T(K) & N & $\rho$ (\AA$^{-3}$) & $E_{\rm Cut}$(Ryd) \\
\hline
OF-AIMD  & 943  &  500  & 0.05290 & 30.25  \\
OF-AIMD  & 1323  &  500  & 0.05071 & 29.25  \\
LRT-CMD  & 943  &  600  & 0.05290 &   \\
LRT-CMD  & 1323  &  500  & 0.05071 &   \\
QHNC-CMD  & 933  &  800  & 0.05331 &   \\
\hline
\end{tabular}
\end{table}

\begin{table}
\caption{Isothermal compresibility $\kappa_T$ (in 
10$^{-11}$ m$^2$ N$^{-1}$),   
self-diffusion coefficient $D$ (in \AA$^2$/ps) and shear viscosity 
coefficient $\eta$ (in GPa$\cdot$ps)   
of liquid Al at the thermodynamic states studied in this work.} 
\label{diffu}
\begin{tabular}{ccccccc}
  & T(K) & $\kappa_T$  & D & $\eta$ \\
\hline
OF-AIMD  & 943  &  2.37  & 0.49 & 1.38 \\
OF-AIMD  & 1323  &  2.38  & 1.05 & 0.85   \\
LRT-CMD  & 943  &  6.57  & 0.58 & 1.24  \\
LRT-CMD  & 1323  &  6.32  & 1.14 & -   \\
QHNC-CMD  & 933  &  7.45  & 0.55 & 1.36   \\
Experiment & 933 &  2.43 \footnotemark[1] &      
&   1.26 \footnotemark[2] \\
\hline
\end{tabular}
\footnotetext[1]{Ref.\  \protect\onlinecite{Seemann}}
\footnotetext[2]{Ref.\  \protect\onlinecite{Shimoji&Itami}}
\end{table}

\end{document}